\documentclass[prd,preprint,superscriptaddress,nofootinbib]{revtex4}
\usepackage{amsmath}
\usepackage{amssymb}
\usepackage{graphicx}
\newcommand{\be}{\begin{equation}}
\newcommand{\ee}{\end{equation}}
\newcommand{\ben}{\begin{eqnarray}}
\newcommand{\een}{\end{eqnarray}}

\begin{document}
\title{Coupled Dark Energy field variation}

\author{Roberto Carlos Garc\'ia-Z\'u\~{n}iga } \affiliation{Facultad de Ciencias, Universidad Aut\'onoma del Estado de M\'exico, Toluca 5000, Instituto literario 100, Edo. Mex.,M\'exico.}

\author{Germ\'{a}n Izquierdo \footnote{ E-mail address:
gizquierdos@uaemex.mx}}\affiliation{Facultad de Ciencias, Universidad Aut\'onoma del Estado de M\'exico, Toluca 5000, Instituto literario 100, Edo. Mex.,M\'exico.}

\begin{abstract}
The variation of the dark energy field is found under the assumption that the dark energy is parametric and interacts with the cold dark matter. Considering that the variation of the field could not exceed the Planck mass, we obtain bounds on the coupling and adiabatic coefficients. Three parameterizations of the adiabatic coefficients are considered and two coupling terms where the energy flows from dark energy to dark matter, or the other way around.
\end{abstract}

 \maketitle

\section{Introduction}
Observational data indicates that the Universe is experimenting an accelerating expansion stage \cite{WMAP, WMAP2, PLANK}. The $\Lambda$CDM model is the most favored to explain the observations: a flat homogeneous universe with three dominant sources of energy density (barionic matter, cold dark matter (CDM) and a cosmological constant $\Lambda$) \cite{rev}. But the $\Lambda$CDM model presents the coincidence problem: the CDM density is of the same order of magnitude of the cosmological constant today. In an attempt to solve this problem, other models have been proposed \cite{rev}. In quintessence and phantom models, the role of the cosmological constant is played by the dark energy (DE), a perfect fluid with constant adiabatic coefficient $w=\rho_{\phi}/p_{\phi}$,  where $-1 \gtrsim w$ and $w \lesssim -1$, respectively. For those models the dark energy density is a dynamical function that evolves with time, and the coincidence problem is transformed in the fine tuning problem: the initial dark energy density and $w$ parameter must take very specific values to explain the observations. Even when assuming the adiabatic coefficient is a dynamical quantity (e.g., DE as a Chapligyn gas, as fluid with parametric $w$, etc.\cite{rev,rev2,rev3}), a fine tuning of the constants appearing in the model is needed to adjust the data. Alternative models are also proposed in the literature to give explanation to observed late acceleration without the need of DE: modified gravity models, backreaction from cosmological perturbations, etc. \cite{rev,other,other1,other2}.

In order to solve (or alleviate) the fine tuning problem of the DE models, the coupled dark energy (CDE) is introduced \cite{rapidc,rapidc1,rapidc2,rapidc3,active,active1,active2,active3,active4,active5,active6,active7,active8,active9,active10,active11,active12,active13,active14,active15,active16}. If the dark energy interacts with the dark matter, they form a dynamical system and can evolve naturally to show similar values today. Given that both sources are the "dark sector" of the Universe (in the sense that no direct observation of any of them have been recorded to date), a possible coupling between them must be not discarded beforehand. CDE models present a late evolution that differs from that of the $\Lambda$CDM model. Perturbations of the background metric evolve in a different way and should be addressed in order to test the validity of the CDE models  and/or to bound their free parameters. An example of this procedure can be found in \cite{ol,ol2,March,Maar,Gavela}, where the density perturbations evolution is studied in different CDE models leading to limits on the coupling parameters of the models. In \cite{He}, density perturbations evolution are also addressed in CDE models when the CDM perturbation experiments a collapse and clusters. The authors demonstrate that the energy density do not always fully cluster with the CDM, and that the cluster abundance count bounds the parameters of the CDE model considered. In \cite{MaLuisa}, the evolution of tensorial perturbations (primordial gravitational waves) in some CDE models is studied.

Bounds on $w$ have been derived  from the reasonable requirement that the variation experienced by the dark
energy scalar field $|\Delta \phi|$ (regardless it may be phantom or quintessence) from any redshift, $z$, within the classical
expansion era, till now should not exceed Planck's mass (see \cite{huang,saridakis,iz10}). This bound looks a rather natural condition and persuasive arguments  have been advanced in its favor
\cite{banks,huang1,yun-gui}. In particular, as noted by Bean {\it et al.} \cite{rachelbean}, the
fractional density of dark energy cannot exceed $5 \%$ and $39 \%$
at the primeval nucleosynthesis and recombination epochs, respectively. Imposing $|\Delta \phi|<M_P$
on every dark energy field would translate on rather loose constraints on
$w$. Given that the DE cannot be observed directly, theoretical bounds and/or indirect measurements are our only tools to understand the nature of it. In \cite{iz10}, the authors find bounds on the adiabatic and the coupling coefficients for CDE from the assumption that the variation of the field cannot exceed the Planck mass focused on two couplings where the energy flows from CDE to CDM. In this work, we proceed in a similar way but considering more general couplings where the energy flows from CDE to CDM as well as in the opposite direction. In this sense, this article generalizes the findings of \cite{iz10} assuming that the parameters can take a wider range of values of physical interest.

The plan of the article is the following. In section \ref{s2}, we present the CDE models as well as the condition over the shift of the field to be computed. In section \ref{s3}, we consider a coupling term proportional to the CDM energy density and we find bounds over the parameters for different choices of the adiabatic coefficient. In section \ref{s4}, we proceed as in the previous section for a coupling term proportional to the CDE density. Finally, in section \ref{s5}, we summarize the findings.

From now on, we assume that a zero subindex refers to the current value of the corresponding quantity; likewise we normalize the scale factor of the metric by setting $a_{0} = 1$.

\section{Coupled Dark Energy field}
\label{s2}
In an attempt to solve the coincidence problem, Coupled Dark Energy (CDE) models are proposed \cite{rapidc,active,review} . Those models assume that the universe is a flat FLRW universe with metric
\[
ds^{2}=-dt^{2}+a(t)^{2}\left[ dr^{2}+r^{2}d\Omega ^{2}\right],
\]%
whose late expansion is dominated by a mixture of three energy density sources: Barionic matter $\rho_{b}$, CDM $\rho_{m}$ and CDE $\rho_{\phi}$. CDE interacts with CDM through the interaction term $Q$ and, consequently, the energy densities evolve as
\begin{eqnarray}
\dot{\rho}_{b}\, &+&\, 3H \rho_{b} = 0 \, , \nonumber \\
\dot{\rho}_{m}\, &+& \, 3 H \rho_{m} = Q \, , \nonumber \\
\dot{\rho}_{\phi}\, &+& \, 3 H (1+w) \rho_{\phi} = - Q \, ,
\label{continuity}
\end{eqnarray}
where $w$ is the CDE adiabatic coefficient $p_{\phi}=w\rho_{\phi}$ and $H$ is the Hubble factor

\be
H^2=\left(\frac{\dot{a}}{a}\right)^2=\frac{1}{3 M_{P}^2}\rho_T, \qquad ( \rho_T=\rho_{b}+\rho_{m}+\rho_{\phi})
\label{H}
\ee

Several forms for the interaction have been proposed in the literature \cite{active}. In this work we will assume the coupling $Q$ is a function of $H \rho_{m}$ and/or $H \rho_{\phi}$. We will consider for simplicity two different coupling terms
\ben
Q_1 = 3\alpha H \rho_{m}\, , \\ Q_2 = 3\alpha H \rho_{\phi} \, ,
\label{int}
\een
where $\alpha$ is an adimensional coupling constant.

In \cite{db}, the authors state that the second law of thermodynamics regarding the entropy of the field gets violated if $\alpha<0$ and the CDE is an effective field, while the entropy is null for a scalar field in a pure quantum state. Assuming that $\alpha>0$, the coupling parameter must be smaller than $0.1$ in order to reproduce the observed values of BAO and CMB anisotropy \cite{ol,ol2}.

In \cite{Maar,Gavela}, the evolution of the linear perturbations of both CDM and CDE are considered for the coupling terms $Q_{1,2}$ concluding that when $\alpha>0$ and $w$ is constant, early non-adiabatic large-scale instabilities are present. On the other hand, considering $\alpha>0$ and a non constant adiabatic coefficient $w=w(a)$ could lead to avoid the instabilities. In \cite{March}, constraints on the negative coupling parameter $\alpha$ are found from the Plank measurements of the CMB anisotropies, finding that $\alpha>-0.90$.

In this work we will assume indistinctly positive and negative values of $\alpha$. We will assume, also, that the adiabatic coefficient $w$ can be constant or dynamical, and that can behave as quintessence or phantom.  Those assumptions are made in order to take a general approach to the variation of the CDE field, which is our main motivation, regardless of the entropy of the field or the early instabilities that could appear.

For the CDE scalar field
\ben
\rho_{\phi}&=& \pm (\dot{\phi})^2/2+V(\phi),\nonumber \\
p_{\phi}&=&\pm (\dot{\phi})^2/2-V(\phi),
\een
where the $\pm$ sing of the kinetic term is positive for a quintessence field ($w>-1$) or negative for a phantom field ($w<-1$), respectively. From the relations of above, it is straightforward that
\be
\dot{\phi}=\sqrt{\pm(\rho_{\phi}+p_{\phi})}=\sqrt{|1+w|\rho_{\phi}}.
\ee
The variation of the CDE field from any instant of the past $t$ until now is
\begin{equation}
\frac{|\Delta \phi|}{M_{P}} = \frac{1}{M_{P}}\int_{t}^{t_{0}} {\dot{\phi}dt} = \int_{a(t)}^{1}{\frac{\sqrt{3|1+w| \Omega_{\phi}(a')}}{a'}}\, da' . \label{Deltaphi}
\end{equation}
where $dt=da'/(a'H(a'))$, $\Omega_{\phi}(a')=3M_{P}\rho_{\phi}(a')/H(a')^2$ and $a_0=1$.

In section \ref{s3} and \ref{s4}, we assume interactions $Q_1$ and $Q_2$, respectively, as well as different functions for the adiabatic coefficient, in order to calculate the variation of the field in each case.

\section{Interaction term proportional to the cold dark matter density}
\label{s3}
By plugging $Q_1$ into (\ref{continuity}), the evolution equations of barionic matter and CDM can be directly solved obtaining
\ben
\rho_{b} &=& \rho_{b0}\, a^{-3}\, , \nonumber \\
\rho_{m} &=& \rho_{m0}\, a^{-3(1- \alpha)} \, , \\
\een
where $\rho_{b0}$, $\rho_{m0}$ are the present day energy density of barionic matter and CDM, respectively. From now on we set energy density units for which $\rho_{T,0}=3 M_P^2 H_0^2=1$. According to recent observations \cite{PLANK}, $\rho_{b0}=0.05$ and $\rho_{m0}=0.27$.

\subsection{Constant $w$}
Assuming $w = w_{0}$, and solving the differential equation for the CDE, we get
\be
\rho_{\phi} = \rho_{\phi0}\, a^{-3(1+w_{0})}\, + \, \rho_{m0} \,
\frac{\alpha}{w_{0}+\alpha} \, \left[ a^{-3(1+w_{0})}-
a^{-3(1-\alpha)} \right],
\label{rhoq1wcte}
\ee
where $\rho_{\phi0}$ is the present day energy density of CDE, $\rho_{\phi0}=0.68$ \cite{PLANK}. If the universe is experimenting an accelerated expansion in the present day (as several observational data suggest),  $\rho_{T,0}+3p_{\phi,0}<0$ and, consequently, $w_0<-1/(3\rho_{\phi0})=-0.49$.

When $\alpha<0$, the solutions of above predict an non-physical negative value of $\rho_{\phi}$ at early times when $w_0<0$, while $\rho_{m}$ is always positive \cite{Maar}\footnote{The reader should note that the coupling parameter $\alpha$ defined in \cite{Maar} is equivalent to $3\alpha$ defined here.}. In other words, $\rho_{\phi}(a_{in})=0$ for $a_{in}^{3(w_0+\alpha)}=(1+\frac{\alpha}{w_{0}+\alpha})$ and negative for $a<a_{in}$.  One possibility to avoid the negative values of $\rho_ {\phi}$ consist in considering that the universe only contains the barionic matter and the CDM for $a<a_{in}$ and that the interaction is somehow 'activated' when the universe reach the scale factor $a=a_{in}$. In this case, it is possible to assume that $\rho_{\phi}$ is given by  (\ref{rhoq1wcte}) for $a>a_{in}$ while it is $0$ for $a<a_{in}$, and that
\be
\rho_m=\left\{
\begin{array}{c}
\rho_{m0}\, {a_{in}}^{3 \alpha}a^{-3}\qquad(a<a_{in}),\\
\rho_{m0}\, a^{-3(1- \alpha)}\qquad(a>a_{in}).\\
\end{array}
\right.
\ee
But then the fine tuning problem the CDE model was trying to solve is back: Why the interaction was null in the past while it is 'activated' at a very concrete instant? With the lack of a deeper study on this phenomenological assumption but having in mind that the scope of the work is to calculate the variation of the field, we chose to check condition $|\Delta \phi|/ M_P<1$ for interaction $Q_1$ and $\alpha<0$ by evaluating integral (\ref{Deltaphi}) from $a_{in}$ until $1$.

When $\alpha$ is positive, both $\rho_{\phi}$ and $\rho_{m}$ are positive defined in the range of the coupling parameter considered, i.e., $\alpha<0.1$. For this choice of parameters, we check condition $|\Delta \phi|/ M_P<1$ by evaluating integral (\ref{Deltaphi}) from the recombination instant (for which $a=1/1090$) until the present day.

\begin{figure}[tbp]
\includegraphics*[scale=0.4]{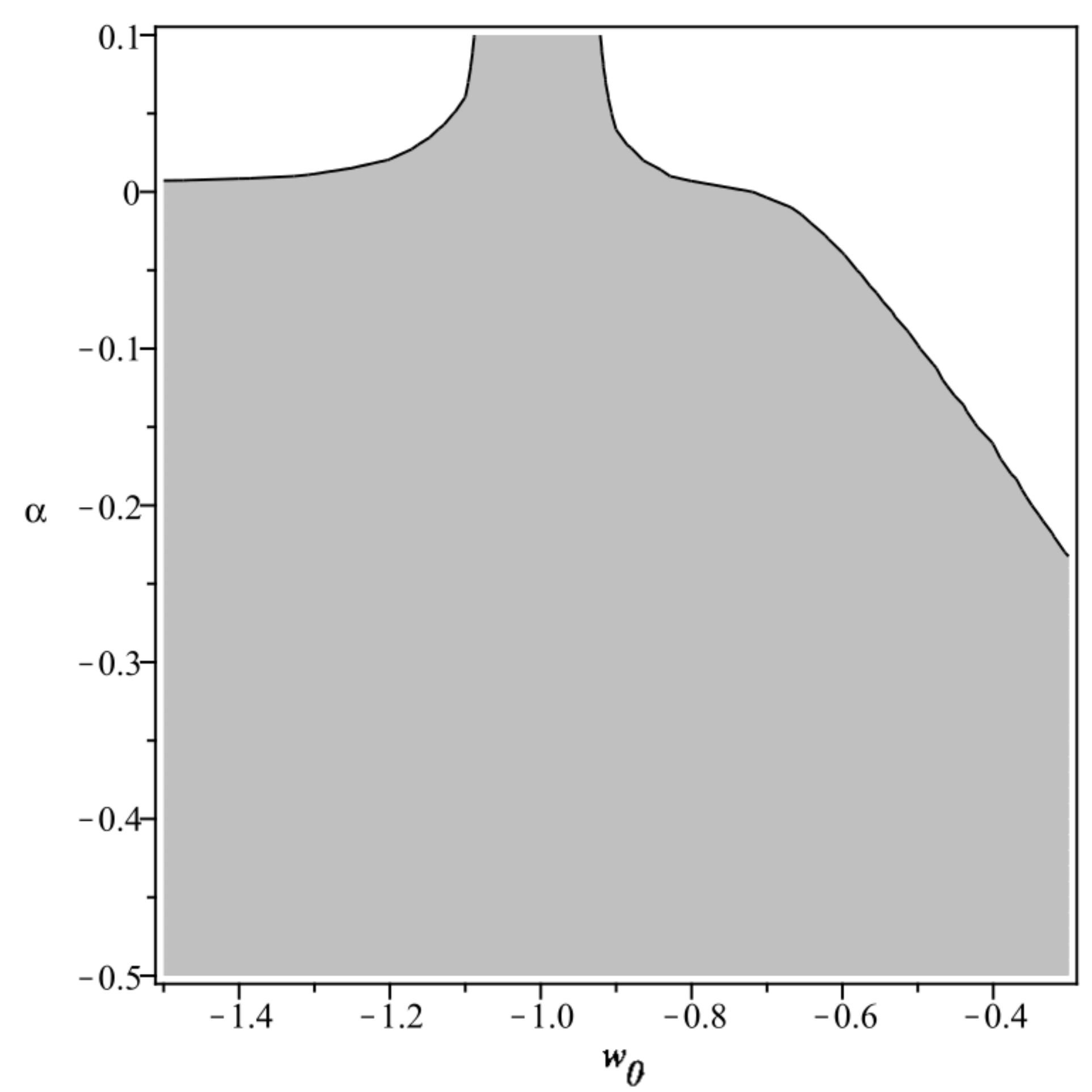}

\caption{Parameters within the shaded region fulfill the condition $|\Delta \phi|/ M_P<1$ for  CDE model with interaction proportional to $\rho_m$ and $w=w_0$.} \label{fig1}
\end{figure}

Fig. \ref{fig1} shows the region of the $w_0$-$\alpha$ space where $|\Delta \phi|/ M_P<1$ in grey. In the quintessence region ($-1<w_0<-0.49$), condition $|\Delta \phi|/ M_P<1$ binds the maximum $w_0$ allowed. The larger the $\alpha$ value is, the more restrictive the bound (e.g., the bound reads $w_0<-0.94$ for $\alpha=0.1$, while reads $w_0<-0.49$ for $\alpha=-0.11$). In fact for $\alpha<-0.11$, no restrictions on the $w_0$  parameter can be found from the variation of the CDE field in the quintessence region. In the phantom region ($w_0<-1$), a lower bound on $w_0$ is found as soon as $\alpha>0$. The variation of the field is always smaller than $M_p$ for negative values of the coupling constant.

\subsection{Linear parametrization}
Except when dark energy is given by the  quantum vacuum, there is
no compelling motivation to assume $w$ constant for the whole
cosmic evolution. In fact, a dynamic $w(a)$ could be used in  models of CDE with $Q_1$ in order to avoid early instabilities \cite{Maar,Gavela}.

The simplest generalization in terms of
redshift ($z=1/a-1$), $w(z) = w_{0} + w_{1} \, z$, is not compatible with
observation for it diverges as $z \rightarrow \infty$. This
prompted the introduction of the more suitable expression $w(z)=
w_{0}\, +\, w_{1}\, \frac{z}{1+z}$ or, equivalently, in terms of
the scale factor \be w(a)= w_{0}\, +\, w_{1}(1-a)
\label{chevallier} \ee by Chevallier and Polarski \cite{ch-p}
(later popularized by Linder \cite{linder}) which does not suffer
from that drawback and behaves nearly linear in $a$. The CDE with adiabatic coefficient given by (\ref{chevallier}) presents different behavior in the past for different choices of the parameters:
\begin{itemize}
\item When $w_{0}+w_{1}> -1$ and $w_{0}> -1$ simultaneously, $w(a)$ is always
larger than $ -1$. The CDE behaves as quintessence at any instant in the past.

\item When $w_{0}+w_{1}< -1$ and $w_{0}< -1$ simultaneously, $w(a)
< -1$ for $0\leq a \leq 1$. The CDE behaves as phantom dark energy during the whole history of the universe.

\item When $w_{0}+w_{1}< -1$ and $w_{0}>-1$, the CDE was phantom in the distant past while behaves as quintessence at the present day. The scale factor of the phantom crossing can be evaluated as $a=1-(1+w_0)/w_1$. We recall this region as mixed region I.

\item When $w_{0}+w_{1}> -1$ and $w_{0}< -1$, the CDE was quintessence in the past while behaves as phantom at the present day. We recall this region as mixed region II.

\item When $w_0>-0.49$, the CDE would predict a decelerated expansion today.

\item Choosing the parameters in a way that $w_{0}+w_{1} > 0$  would lead to a dark energy dominance in the distant past which can interfere with the structure formation scenario, the radiation dominated epoch and/or the nucleosynthesis.  We recall this region as forbidden, as would clearly be in contradiction with observational data.

\end{itemize}

In this case, the expression for $\rho_{\phi}$ must be found numerically. When $\alpha<0$, the solutions also predict an non-physical negative value of $\rho_{\phi}$ at early times (as in the $w$ constant case), while $\rho_{m}$ is always positive. For every choice of $w_0$, $w_1$ and $\alpha<0$, it is possible to compute $a_{in}$ in the range $[0,1]$ for which $\rho_{\phi}(a_{in})=0$ and, then, check condition $|\Delta \phi|/ M_P<1$ by evaluating integral (\ref{Deltaphi}) from $a_{in}$ until $1$. When $\alpha$ is positive, both $\rho_{\phi}$ and $\rho_{m}$ are positive defined in the past. Thus, we check condition $|\Delta \phi|/ M_P<1$ by evaluating integral (\ref{Deltaphi}) from the recombination scale factor (for which $a=1/1090$) until the present day.

\begin{figure}[tbp]
\includegraphics*[scale=0.4]{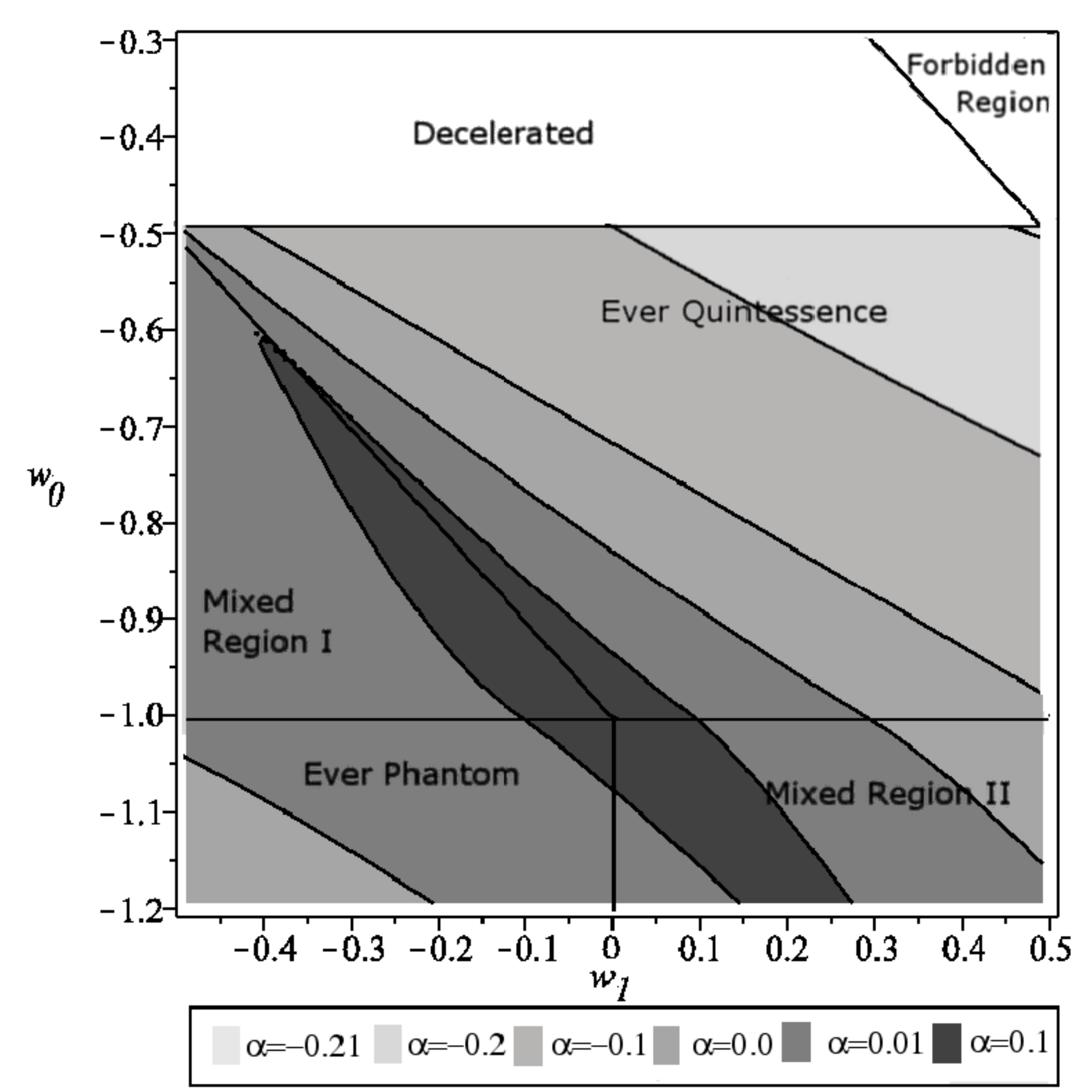}

\caption{Variation of the field $|\Delta \phi|$ for  CDE model with interaction proportional to $\rho_m$ and $w(a)=w_0+w_1(1-a)$. Choosing $w_0$ and $w_1$ within the corresponding shaded $\alpha$ contour leads to $|\Delta \phi|<M_P$. For the numerical work, we take $w_0 $ in the interval $[-0.49,-1.20]$ and $w_1$ in the interval $[-0.50,0.50]$ and divide both intervals in steps of $0.01$.} \label{fig2}
\end{figure}

Fig. \ref{fig2} shows the region of the $w_1$-$w_0$ space where $|\Delta \phi|/ M_P<1$ as the shaded surfaces for different choices of $\alpha$ parameter. For $\alpha=0.1$, the area of the $w_0$-$w_1$ parameter space that fulfills $|\Delta \phi|/ M_P<1$ is restricted to the darker shaded region. The lower the value of $\alpha$, the more wider the region of allowed parameters is. For $\alpha=0$, only some choices of the quintessence region are restricted. These findings are in good agrement with the results reported in \cite{iz10}, although the present day energy densities used in this work are $\rho_{b0}=0.05$, $\rho_{m0}=0.27$ and $\rho_{\phi0}=0.68$ instead of the ones used in \cite{iz10} ( $0.04$, $0.24$ and $0.72$, respectively).

For negative values of $\alpha$, the restriction over the parameters in the quintessence region is reduced. For $\alpha\leq-0.21$ no restrictions over the parameters $w_0$ and $w_1$ can be found from the variation of the field.

\subsection{Barboza-Alcaniz parametrization}
As readily noted, Chevallier-Polarski-Linder's parametrization
(\ref{chevallier}) implies that $w(a)$ diverges as $a \rightarrow
\infty$ (i.e., in the far future). To avoid this unpleasant feature
Barboza and Alcaniz proposed  the ansatz $w (z) = w_{0} \,+ \,
w_{1} \, \frac{z(1+z)}{1\, +\, z^{2}}$ or, equivalently,
\be
w(a)=w_{0} \, +\, w_{1} \, \frac{1-a}{1-2a\, +\, 2a^{2}} \, ,
\label{alcaniz}
\ee
which ensures that $w(a)$ stays bounded in the
whole interval $0 \leq a < \infty$ \cite{b-a}. We will denote it as the BA parametrization. The CDE with adiabatic coefficient given by (\ref{alcaniz}) presents similar regions as the considered in the previous parametrization.

Assuming this novel parametrization alongside the interaction term
$Q_1$, $\rho_{\phi}$ must be calculated numerically. Again, choosing $\alpha<0$ leads to negative $\rho_{\phi}$ at early times, while choosing $\alpha>0$ does not. As in the previous cases, we compute $a_{in}$ (while necessary) for every choice of $w_0$, $w_1$ and $\alpha$, and we evaluate (\ref{Deltaphi}) from $a_{in}$ up to $a=1$ (or, from $a=1/1090$ otherwise).

\begin{figure}[tbp]
\includegraphics*[scale=0.4]{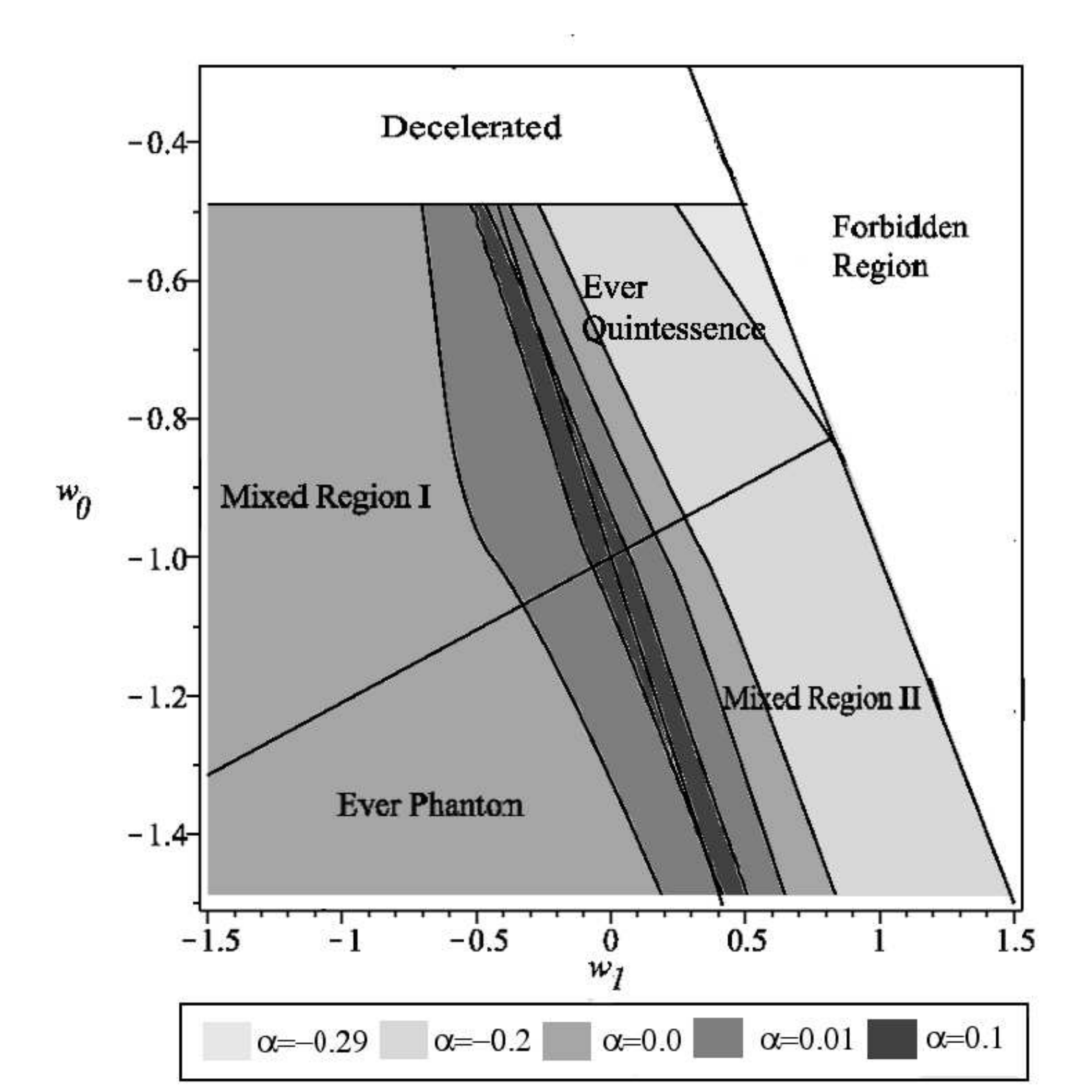}

\caption{Variation of the field $|\Delta \phi|$ for  CDE model with interaction proportional to $\rho_m$ and $w(a)=w_{0}+w_{1} \frac{1-a}{1-2a+ 2a^{2}}$. The shaded contours fulfills the condition $|\Delta \phi|<M_P$ for different choices of $\alpha$. For the numerical work, we take $w_0 $ in the interval $[-0.49,-1.50]$ and $w_1$ in the interval $[-1.50,1.50]$ and divide both intervals in steps of $0.01$.} \label{fig3}
\end{figure}

Figure \ref{fig3} illustrates the regions of the the $w_1$-$w_0$ space for which $|\Delta \phi|<M_P$ for some choices of $\alpha$. The results for $\alpha=0.1$, $0.01$ and $0.0$ are very similar to those reported in the literature \cite{iz10} with small differences related to the different initial conditions considered. The restriction for $\alpha<0$ only affects a part of the ever quintessence and mixed region II parameter space. The smaller the $\alpha$, the thinner the forbidden region. For $\alpha<-0.2$, all the parameters in the mixed region II lead to $|\Delta \phi|<M_P$, and only some choices of the ever quintessence region are discarded. Finally, for $\alpha<-0.29$, $|\Delta \phi|<M_P$ no matter what choice of $w_0$ and $w_1$ is done.

\section{Interaction term proportional to the coupled dark energy density}
\label{s4}
In this section we take up $Q_2= 3 \, \alpha H \rho_{\phi}$ for the
coupling  term alongside different expressions for the equation of state
parameter. The evolution equations of barionic matter lead again to $\rho_b=\rho_{b0}a^{-3}$, while both $\rho_{m}$ and $\rho_{\phi}$ must be solved in a case by case basis.

\subsection{Constant $w$}
By plugging $Q_2$ into (\ref{continuity}),
assuming $w = w_{0}$, and solving the differential equations for the CDE and CDM, we get
\ben
\rho_{m}&=& \rho_{m0}\, a^{-3}+\frac{\alpha}{w_0+\alpha}\,
\rho_{\phi0} \,
a^{-3}\left[1-a^{-3(w_0+\alpha)}\right] \, , \nonumber \\
\rho_{\phi}&=&\rho_{\phi0} \, a^{-3(1+w_0+\alpha)} \, .\\
\label{rhoq2wcte}
\een

For this interaction, neither  $\rho_{\phi}$ nor $\rho_{m}$ are negative at early times \cite{Maar,Gavela}. We check condition $|\Delta \phi|/ M_P<1$ by evaluating integral (\ref{Deltaphi}) from the recombination scale factor until the present day.

\begin{figure}[tbp]
\includegraphics*[scale=0.4]{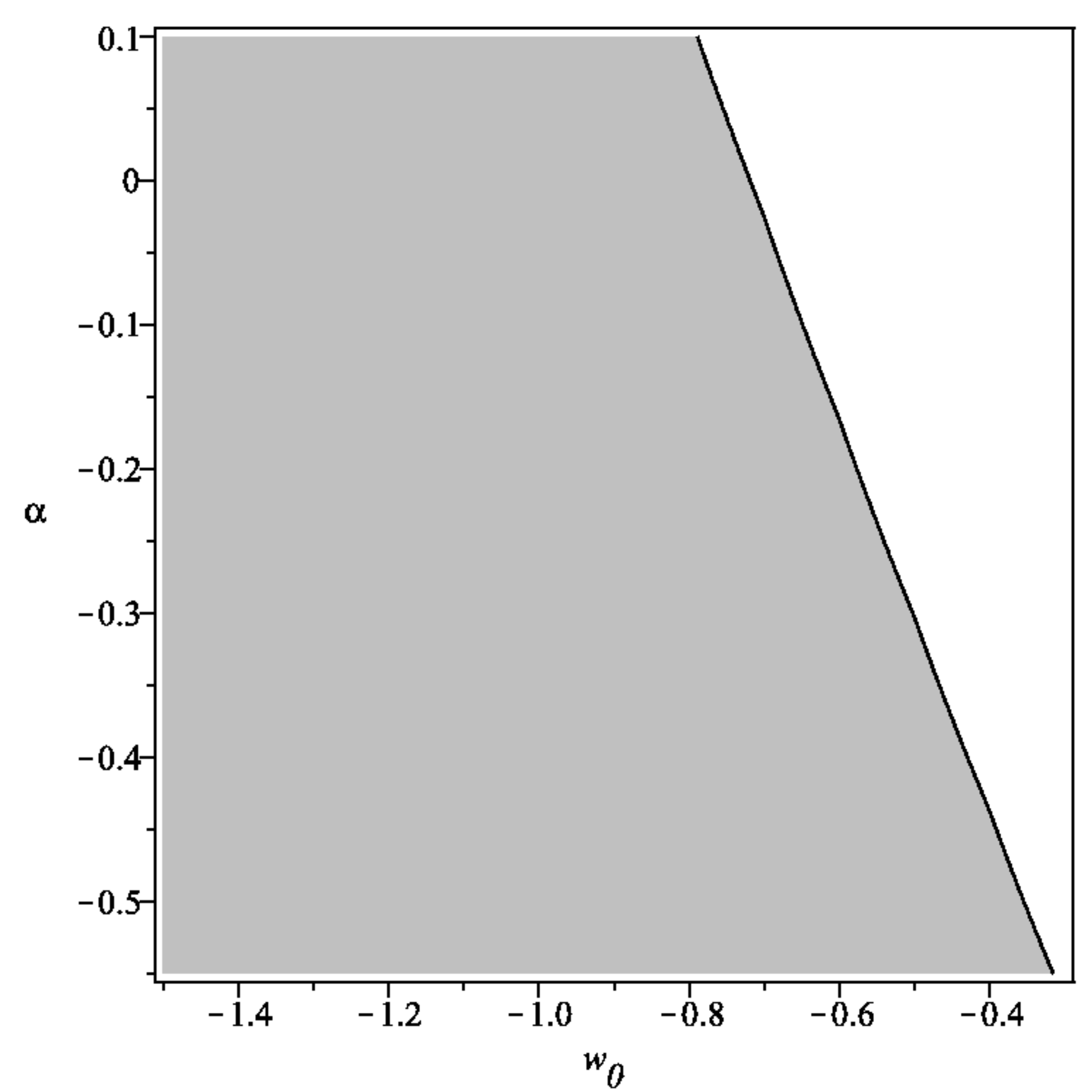}

\caption{Condition $|\Delta \phi|/ M_P<1$ for  CDE model with interaction proportional to $\rho_{\phi}$ and $w=w_0$ is fulfilled if the parameters are chosen within the shaded region.} \label{fig4}
\end{figure}

Fig. \ref{fig4} shows the region of the $w_0$-$\alpha$ space where $|\Delta \phi|/ M_P<1$ in grey. In the quintessence region ($-1<w_0<-0.49$), condition $|\Delta \phi|/ M_P<1$ binds the maximum value of $w_0$ allowed. The smaller the $\alpha$ value is, the less restrictive the bound. In fact for $\alpha<-0.32$ no restrictions on the $w_0$  parameter can be found from the variation of the CDE field in the quintessence region. In the phantom region ($w_0<-1$), no lower bound on $w_0$ is found, no matter $\alpha>0$ (something already reported in \cite{iz10}) or $\alpha<0$.

\subsection{Linear parametrization}

In this case, we can analytically find $\rho_{\phi}$ and  $\rho_m$ as

\begin{figure}[tbp]
\includegraphics*[scale=0.4]{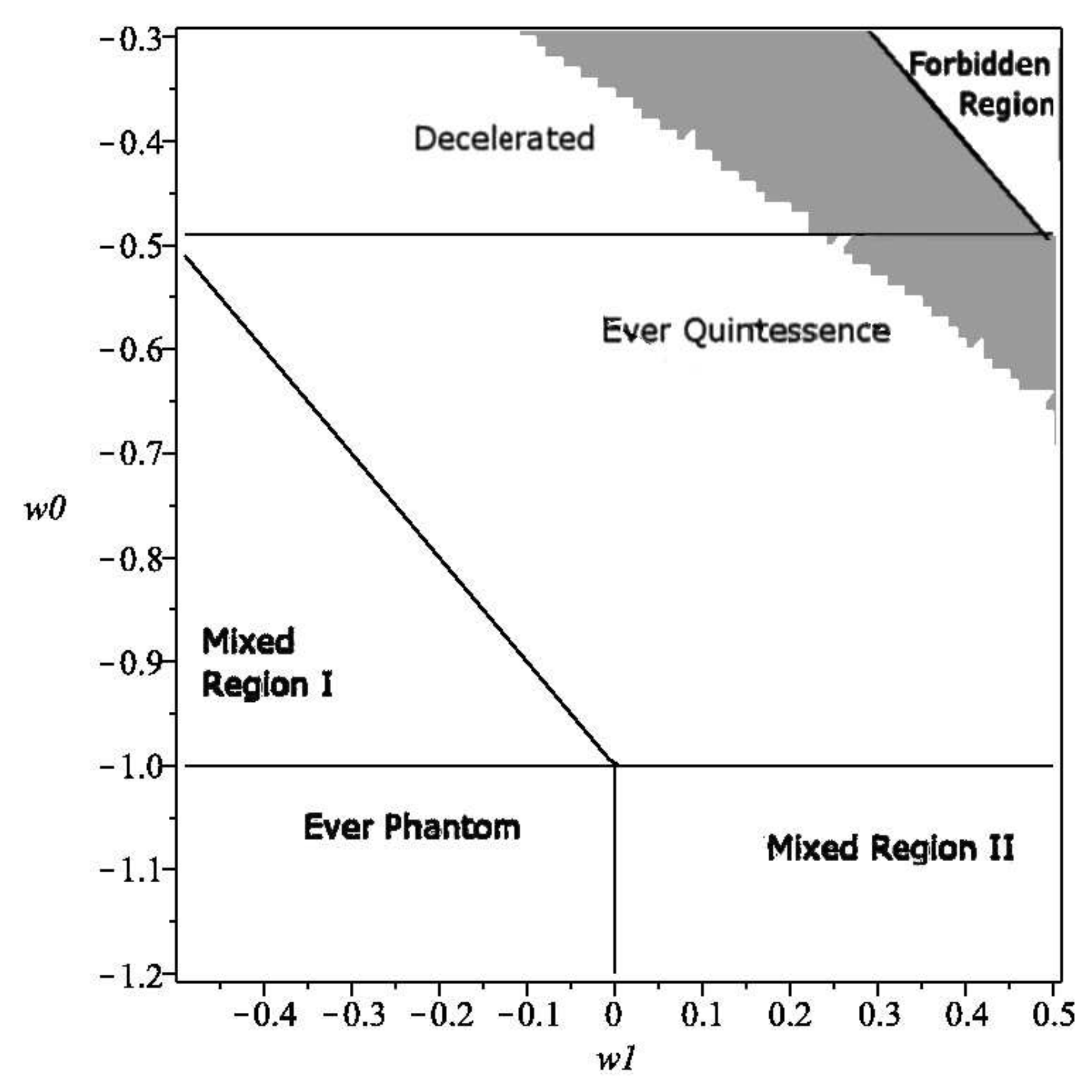}
\includegraphics*[scale=0.4]{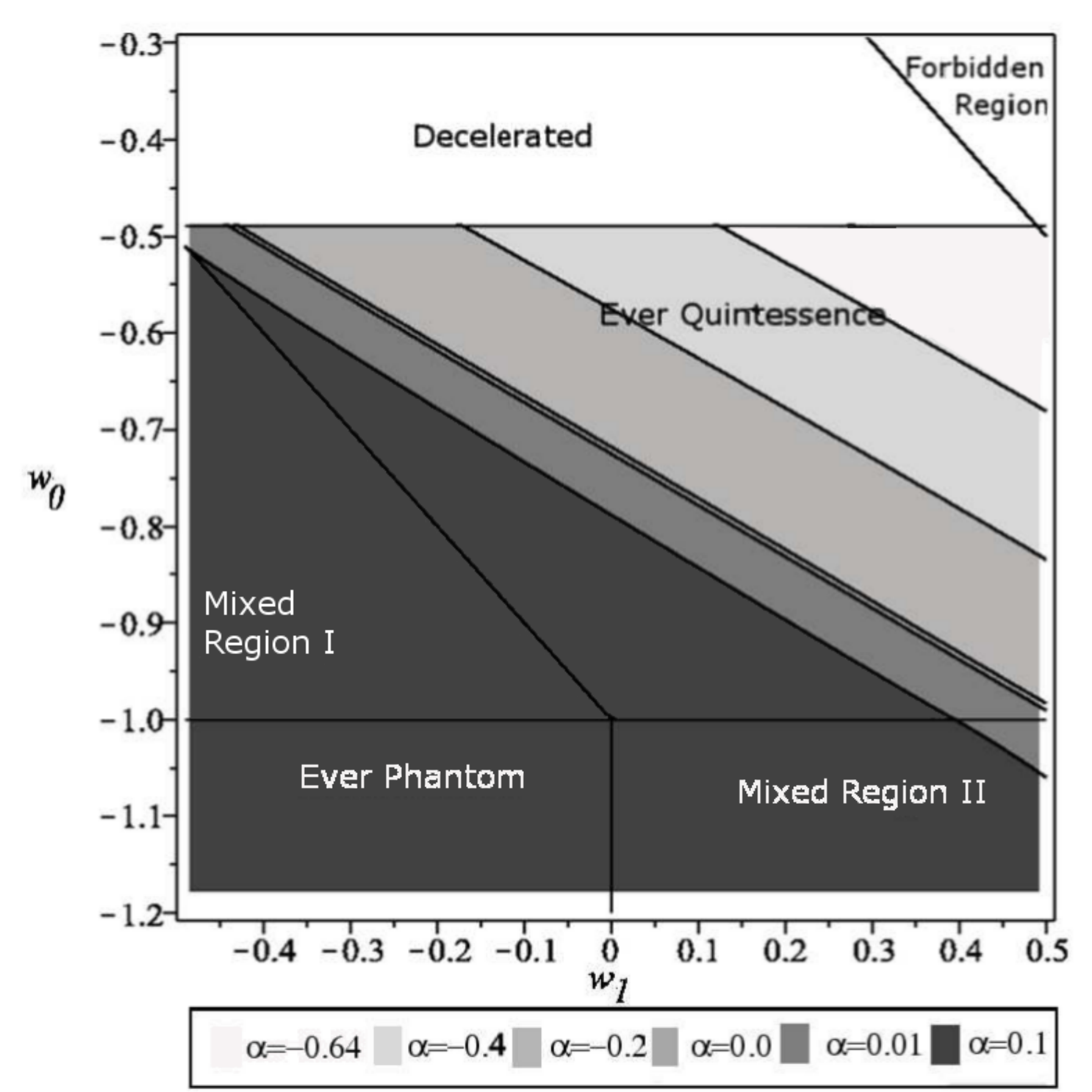}

\caption{a) Shaded region represents the choices of $w_0$ and $w_1$ that lead to early negative $\rho_m$ when $\alpha=0.1$. The irregular shape of the contour is a consequence of the partition of the $w_0$ and $w_1$ intervals done for the numerical work.  b)Variation of the field $|\Delta \phi|$ for  CDE model with interaction proportional to $\rho_m$ and $w(a)=w_0+w_1(1-a)$. The shaded contours fulfills the condition $|\Delta \phi|<M_P$ for different choices of $\alpha$.} \label{fig5}
\end{figure}

\ben
\rho_{\phi}&=&\rho_{\phi0}e^{-3w_1(1-a)}a^{-3(1+ W)},\\
\rho_m &=& \rho_{m 0}a^{-3}+3 \alpha\rho_{\phi 0}{a}^{-3}e^{-3\,w_1}\left( -3\,w_1 \right) ^{3\,W}
\left[-\,\,\Gamma  \left( -3\,W,-3\,w_1\,a \right)  \,\, +\,\,\Gamma  \left( -3\,W,-3\,w_1 \right)\right],\nonumber
\een
where $W=w_0+w_1+\alpha$ and $\Gamma(i,j)$ is the incomplete Gamma function. From the evolution equations of above, it is clear that $\rho_{\phi}$ is always positive in the past while $\rho_m$ can take negative values. When $\alpha=0.1$, there is a region on the $w_0$-$w_1$ space that leads to negative early values of $\rho_m$, something already reported in \cite{iz10}. Fig. \ref{fig5}a shows the region of early  negative $\rho_m$ when $\alpha=0.1$. For $\alpha \leq 0.01$, all choices of $w_0$-$w_1$ in the allowed region lead to an always positive $\rho_m$.

In this case, we check condition $|\Delta \phi|/ M_P<1$ by evaluating integral (\ref{Deltaphi}) from the recombination scale factor until the present day, avoiding the non physical region for $\alpha=0.1$.

Fig. \ref{fig5}b shows the region of the $w_1$-$w_0$ space where $|\Delta \phi|/ M_P<1$ as the shaded surfaces for different choices of $\alpha$ parameter.

For $\alpha\leq -0.64$, no bounds can be found over the parameters $w_0$ and $w_1$ from the condition $|\Delta \phi|/ M_P<1$.

\subsection{Barboza-Alcaniz parametrization}

\begin{figure}[tbp]
\includegraphics*[scale=0.4]{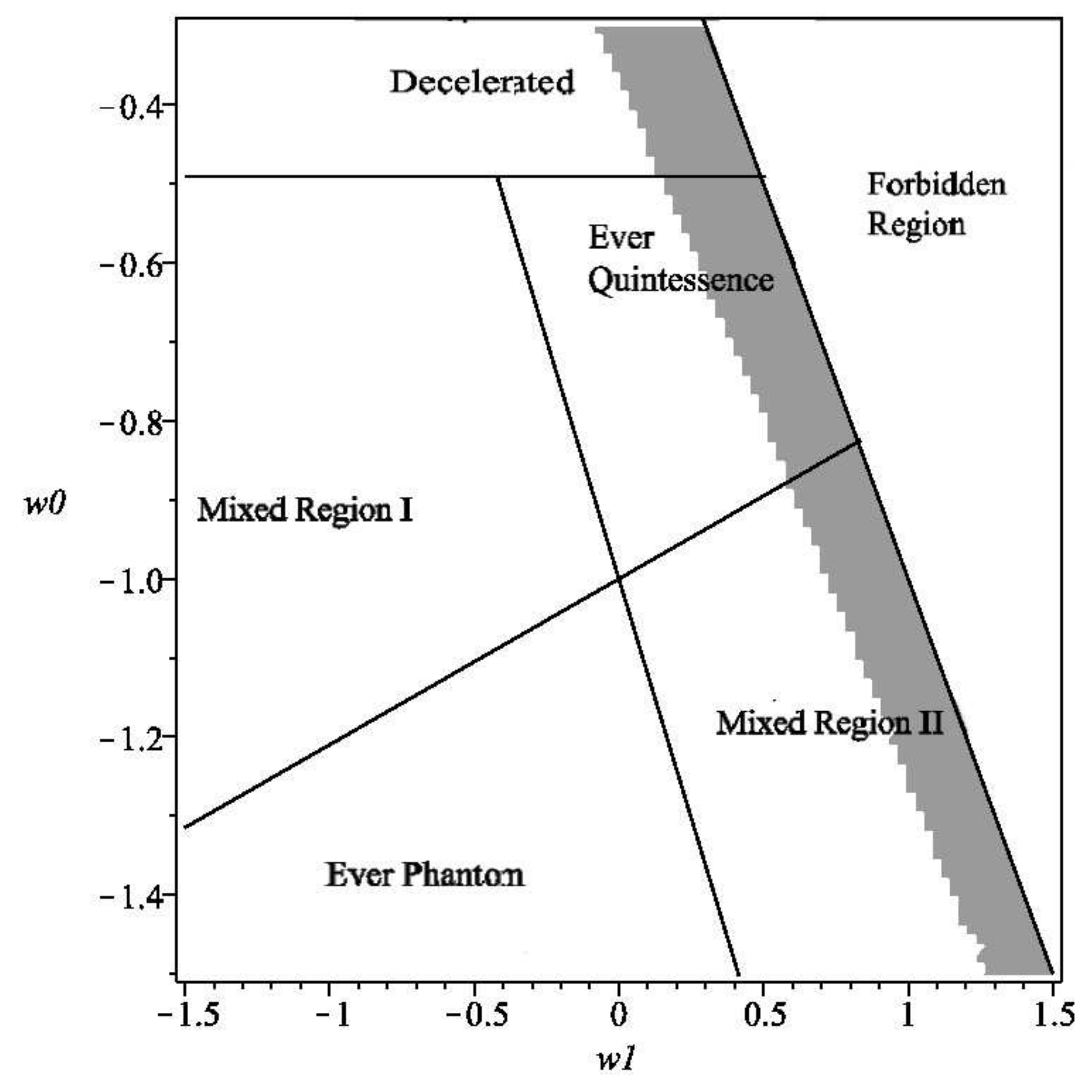}
\includegraphics*[scale=0.4]{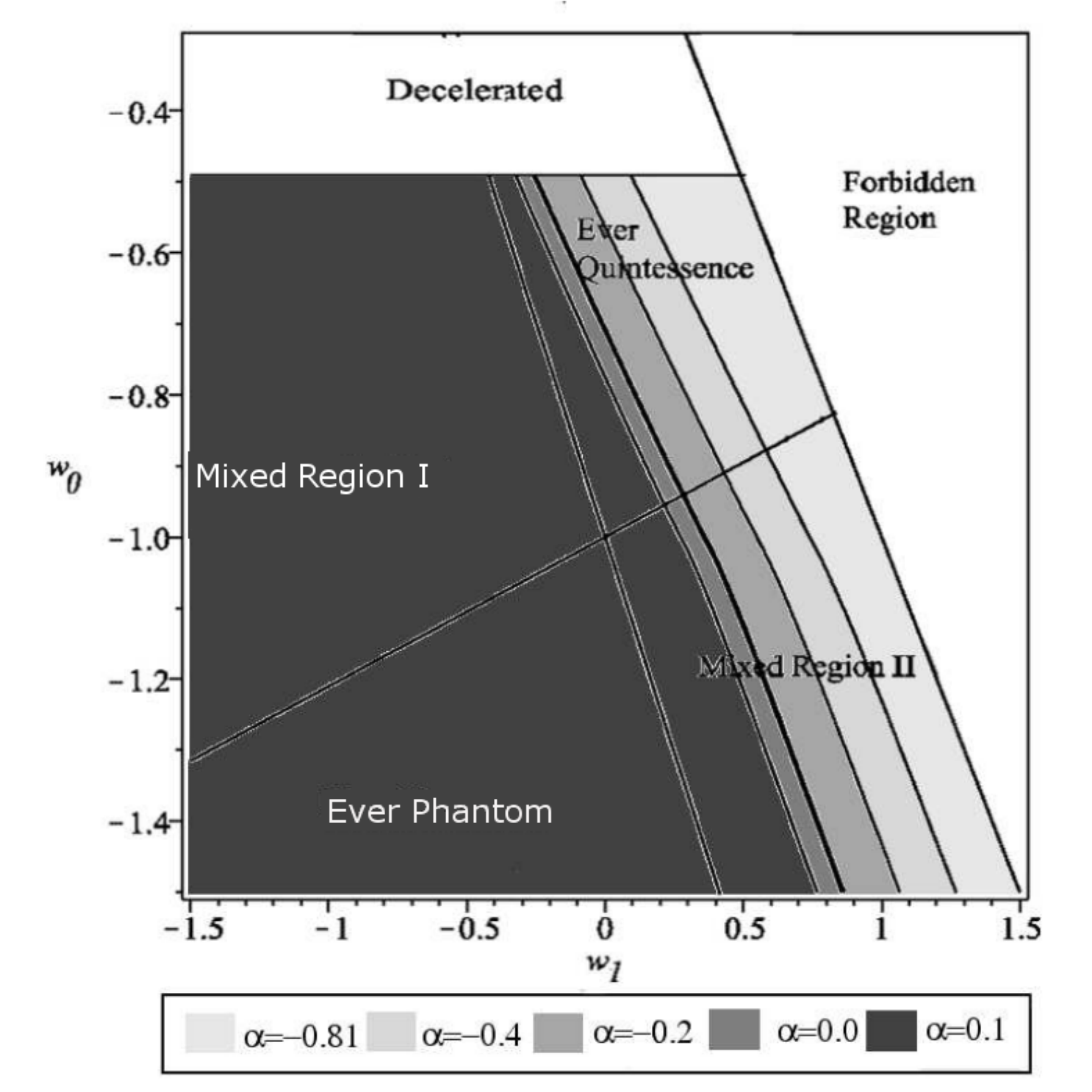}
\caption{a) Shaded region represents the choices of $w_0$ and $w_1$ that lead to early negative $\rho_m$ when $\alpha=0.1$. The irregular shape of the contour is a consequence of the partition of the $w_0$ and $w_1$ intervals done for the numerical work. b)Variation of the field $|\Delta \phi|/M_P$ for  CDE model with interaction proportional to $\rho_{\phi}$ and $w(a)=w_{0}+w_{1} \frac{1-a}{1-2a+ 2a^{2}}$. The shaded contours fulfills the condition $|\Delta \phi|<M_P$ for different choices of $\alpha$.} \label{fig6}
\end{figure}

Assuming this novel parametrization alongside the interaction term
$Q_2$, $\rho_{\phi}$ reads
\be
\rho_{\phi}=\rho_{\phi 0} a^{-3(1+W)} \left( 1-2\,a+2\,{a}^{2} \right) ^{3 w_1/2},
\ee
where $W=w_0+w_1+\alpha$ and which is always positive. The CDM density must be computed numerically. Again, $\rho_m$ is negative at early times when $\alpha=0.1$ and $w_1$-$w_0$ are chosen in the shaded region of Fig. \ref{fig6}a.

Fig. \ref{fig6}b illustrates the regions of the the $w_1$-$w_0$ space for which $|\Delta \phi|<M_P$ for some choices of $\alpha$. The results for $\alpha=0.1$ and $0.0$ are very similar to those reported in the literature \cite{iz10} with small differences related to the different initial conditions considered. The restriction for $\alpha<0$ reduces to a small contour in the quintessence and mixed II regions. For $\alpha<-0.81$, $|\Delta \phi|<M_P$ no matter what choice of $w_0$ and $w_1$ is done in the allowed region.

\section{Conclusions}
\label{s5}
Motivated by the reasonable assumption that the variation $\Delta
\phi$ of the field driving the present phase of cosmic accelerated
expansion should not exceed Planck's mass we have numerically
calculated the said evolution since last scattering till now for
different expressions of the equation of state parameter $w(a)$
and two couplings (the coupling strength of them parameterized by $\alpha$).  This constrains the parameter space $(w_{1}$-$w_{0})$ as shown in the six cases studied and represented in Figs. \ref{fig1}-\ref{fig6}.

When $\alpha>0$ (CDE energy density transferred to the CDM) and considering interaction $Q_1$, the condition puts a higher and a lower bound in the $w_0$ considered for constant adiabatic coefficient. For linear and BA parameterizations, this bound restrict parameters $w_0$ and $w_1$. Those bounds are more relaxed when a smaller $\alpha$ is considered. In fact, for $\alpha=0$, it only affects the region of quintessence in the constant parametrization (the quintessence and mixed II regions for the other parameterizations), making possible any choice of parameters in the phantom (and mixed I region). For $\alpha>0$ and $Q_2$, limits are found in the quintessence (and mixed II) region while no bounds are found in the phantom (and mixed I) region. In this case when $\alpha=0.1$ and for some choice of parameters, $\rho_m$ evolves to negative values in the past. Those results are in good agreement with the results reported in \cite{iz10}, with small differences related to the different present day values of the energy densities considered.

When $\alpha<0$ (flux of energy from CDM to CDE) and considering interaction proportional to $\rho_m$, $Q_1$,  the variation of the field is computed from an initial scale factor $a_{in}$ (for which $\rho_{\phi}(a_{in})=0$) up to today, $a_0=1$. In this case, $|\Delta \phi|<M_P$ puts a limit over the adiabatic coefficient in the quintessence region for constant $w$ (and mixed II regions also, for the other two parameterizations considered). The phantom (and mixed I) region always lead to a variation of the field lower than the Planck's mass. For $\alpha<0$ and $Q_2$, both $\rho_{\phi}$ and $\rho_m$ are always positive defined and the integral is computed from the recombination scale factor up to now. In the latter case, similar bounds are found as in the former case.

In all cases there exist a negative value of $\alpha$ for which condition $|\Delta \phi|<M_P$ is always fulfilled, no matter the value of $w_0$ (and $w_1$). This critical value of $\alpha$ depends on the interaction and parametrization chosen: $-0.11$ for interaction $Q_1$ and $w$ constant, $-0.21$ for $Q_1$ and linear parametrization; $-0.29$ for $Q_1$ and BA parametrization; $-0.32$ for interaction $Q_2$ and constant $w$; $-0.64$ for $Q_2$ and linear parametrization; and, finally, $-0.81$ for $Q_2$ and BA parametrization. In this sense, we can conclude that the critical $\alpha$ for interaction $Q_2$ take lower values than the one of interaction $Q_1$, when the same parametrization is considered.



\begin{thebibliography}{99}
\bibitem{WMAP} C.L. Bennett, et.al., Astrophys. Journal S., 208, 20B (2013).
\bibitem{WMAP2}    G.F. Hinshaw, et.al.,  Astrophys. Journal S., 208, 19H (2013).

\bibitem{PLANK}  P. A. R. Ade, et al., arXiv:1303.5076v1.

\bibitem{rev}   E. J. Copeland, M. Sami, S. Tsujikawa, Int.J.Mod.Phys.D \textbf{15} 1753-1936,(2006).
\bibitem{rev2} J. Frieman, M. Turner, D. Huterer,   	 Ann.Rev.Astron.Astrophys. \textbf{46} 385-432 (2008).
\bibitem{rev3} R. R. Caldwell and M. Kamionkowski,	 Ann.Rev.Nucl.Part.Sci. \textbf{59} 397-429, (2009).

\bibitem{other} S. Zorba, Mod. Phys. Lett. A, 27, 1250106 (2012).
\bibitem{other1} R. Sussman, Class. and Quant. Grav., vol 28, pp 235002 (2011). 
\bibitem{other2} S. Del Campo, V. H. Cirdenas, and R. Herrera, Mod. Phys. Lett. A 27, 1250213 (2012).

\bibitem{rapidc}
S. del Campo, R. Herrera, and D. Pav\'{o}n, Phys. Rev. D
\textbf{78}, 021302(R) (2008).
 \bibitem{rapidc1}S. del Campo, R. Herrera, and D. Pav\'{o}n, JCAP 01(2009) 020.
 \bibitem{rapidc2} X. Chen, Y. Gong, E. N. Saridakis, JCAP 0904 (2009) 001.
\bibitem{rapidc3}X. Chen, Y. Gong, E. N. Saridakis, Int.J.Theor.Phys. \textbf{53} (2014) 469-481.

\bibitem{active}
C. Wetterich, Nucl.Phys. B \textbf{302}, 668 (1988).
\bibitem{active1} W. Zimdahl,D. Pav\'{o}n, and L.P. Chimento, Phys. Lett. B \textbf{521}, 133
(2001).
\bibitem{active2} L. Amendola and D. Tocchini-Valentini, Phys. Rev. D
\textbf{66}, 043528 (2002).
\bibitem{active3} G. Farrar and P.J.E. Peebles,
Astrophys. J \textbf{604}, 1 (2004).
\bibitem{active4} H. Zhang and Z.-H. Zhu, Phys. Rev. D \textbf{73},
043518 (2006).
\bibitem{active5} S. del Campo, R. Herrera, and D. Pav\'{o}n, Phys.
Rev. D \textbf{74}, 023501 (2006).
\bibitem{active6}  R. Manini and S. Bonometto,
JCAP 06(2007)020.
\bibitem{active7} Z.-K. Guo, N. Ohta, and S. Tsujikawa, Phys. Rev.
D. \textbf{76}, 023508 (2007).
\bibitem{active8} J.H. He and B. Wang, JCAP
06(2008)010.
\bibitem{active9} T. Koivisto, and D. Mota, Astrophys. J. \textbf{679},
1 (2008).
\bibitem{active10} X. Fu,  H. Yu, and P. Wu, Phys. Rev. D \textbf{78},
063001 (2008).
\bibitem{active11}P.M. Sutter and P.M. Ricker, Astrophys. J.
\textbf{687}, 7 (2008).
\bibitem{active12} J. Valiviita, E. Majerotto, and R.
Maartens, JCAP 07(2008)020.
\bibitem{active13} J.-H. He, B. Wang, and  E. Abdalla,
Phys. Lett. B \textbf{671}, 139 (2009).
\bibitem{active14}J.-H. He, B. Wang, and
Y.P. Jing, JCAP 07 (2009) 030.
\bibitem{active15}C. E. Pellicer et al.,  Mod. Phys. Lett. A 27, 1250144 (2012). 
\bibitem{active16}F. R. Klinkhamer, Mod. Phys. Lett. A 27, 1250150 (2012).

\bibitem{ol} G. Olivares, F. Atrio-Barandela, D. Pav\'on, Phys.Rev. D \textbf{71}  063523 (2005).
\bibitem{ol2}G. Olivares, F. Atrio-Barandela, D. Pav\'on, Phys. Rev. D 74 043521 (2006).

\bibitem{March} V. Salvatelli, A. Marchini, L. Lopez-Honorez, and O.
Mena, Phys. Rev. D \textbf{88}, 023531 (2013).

\bibitem{Maar}J. Valiviita, E. Majerotto, R. Maartens, JCAP \textbf{0807} 020, (2008).

\bibitem{Gavela} M.B. Gavela, D. Hernandez, L. Lopez Honorez, O. Mena, S. Rigolin, JCAP \textbf{0907} 034, (2009).

\bibitem{He} Jian-Hua He, Bing Wang, E. Abdalla, D. Pavon, J. of Cosmology and Astroparticle Phys. \textbf{1012} 022 (2010).
\bibitem{MaLuisa}  M. L. Sosa-Almazán and G. Izquierdo Gen. Rel. and Grav. \textbf{46}, Issue 7 (2014).
\bibitem{huang}
Q.-G. Huang, Phys. Rev. D \textbf{77}, 103518 (2008).
\bibitem{saridakis}
E. N. Saridakis, Phys. Lett. B \textbf{676}, 7 (2009).
\bibitem{iz10}
G. Izquierdo
and D. Pav\'on, Physics Letters B \textbf{688} 115 (2010).

\bibitem{banks}
T. Banks, M. Dine, P.J. Fox, and E. Gorbatov, JCAP 06(2003)001.
\bibitem{huang1}
Q.G. Huang, Phys. Rev. D \textbf{76}, 061303(R) (2007).
\bibitem{yun-gui}
X.-M Chen, J. Liu, and Y-G. Gong, Chin. Phys. Lett \textbf{25}, 8
(2008).
\bibitem{rachelbean}
R. Bean, S.H. Hansen, and A. Melchiorri, Phys. Rev. D \textbf{64},
103508 (2001).

\bibitem{review}
F. Atrio-Barandela and D. Pav\'{o}n, ``Interacting dark energy" in
{\em Dark Energy-Current Advances and Ideas}, edited by J.R. Choi
(Research Signpost, Trivandrum, Kerala, India; in press, 2010).
\bibitem{db}
D. Pav\'{o}n and B. Wang, Gen. Relativ. Grav. \textbf{41}, 1
(2009).


\bibitem{ch-p}
 M. Chevallier and D. Polarski, Int. J. Mod. Phys. D \textbf{10},
213 (2001).
\bibitem{linder}
E.V. Linder, Phys. Rev. Lett. \textbf{90}, 091301 (2003).

\bibitem{b-a}
E.M. Barboza Jr. and J.S. Alcaniz, Phys. Lett. B \textbf{666}, 415
(2008).

\end{thebibliography}
\end{document}